\documentstyle[12pt,epsfig]{article} 
\begin{document}
\title{Measurement--induced interference in an inhomogeneous gravitational field.}
\author{ A. Camacho
\thanks{email: acamacho@aip.de} \\
Astrophysikalisches Institut Potsdam. \\
An der Sternwarte 16, D--14482 Potsdam, Germany.}

\date{}
\maketitle

\begin{abstract}

A very interesting quantum mechanical effect is the emergence 
of gravity--induced interference, which has already been detected. 
This effect also shows us that gravity is at the quantum level not a purely 
geometric effect, the mass of the employed particles appears explicitly in the 
interference expression. In this work we will generalize some previous results. 
It will be shown that the introduction of a second order approximation in the propagator of 
a particle, immersed in the Earth's gravitational field, and whose 
coordinates are being con\-ti\-nuous\-ly monitored, allows us to include, in the corresponding complex oscillator, a frequency which 
now depends on the geometry of the source of the gravitational field, a fact that 
is absent in the case of a homogeneous field. Using this propagator we will analyze the interference pattern of two particle 
beams whose 
coordinates are being continuously monitored. We will compare our results againt the 
case of a homogeneous field, and also against the measurement ouputs of the 
Colella, Overhauser, and Werner experiment, and find that the difference in the 
dependence upon the geometry of the source of the gra\-vi\-ta\-tional field could render detectable differences in 
their respective measurement outputs. 

\end{abstract}
\newpage
\section{Introduction.}
\bigskip

Currently a very interesting and also controversial point comprises the role that gra\-vi\-ty plays 
at the quantum level. 
We already know that at this level (here quantum level means the effect 
that a classical gravitational field has on a quantum particle, i.e., no quantum 
theory of gravity is contemplated here) gravity is not a purely geo\-me\-tric effect. Indeed, the 
experimentally detected gravity--induced in\-ter\-fe\-rence pattern emerging from two particle beams 
depends explicity on the mass of the employed particles [1]. This is not the case 
in classical mechanics, in this context mass does not appear in the 
motion equation of a particle trajectory, i.e., gravity in classical mechanics can be 
geometrized. 

An additional fundamental problem in current physics comprises the so called quantum 
measurement problem. In connection with this conceptual difficulty, there are already 
may attempts which try to solve it, one of them is the so called restricted path integral formalism (RPIF) [2]. 
This formalism explains a continuous quantum measurement with the introduction of a restriction on 
the integration domain of the corresponding path integral. This last condition can also be reformulated in terms 
of a weight functional that has to be considered in the path integral. 

This formalism has been employed in several situations, i.e., the analysis of 
the response of a gravitational wave antenna of Weber type [2], 
the measuring process of a gravitational wave in a laser--interferometer [3], 
or even to explain the emergence of the classical concept of time [4]. 
In the context of this model it is also important to add that there are 
already some theoretical predictions that could 
render a feasible framework which could allow us to confront RPIF against 
experimental outputs [5], but it seems also that more work is needed in this direction.

At this precise point we may join RPIF and the emergence of a gravity--induced 
interference. It has already been proved that if the position 
of two particle beams, immersed in a homogeneous gravitational field, is measured continuously, 
then a new testing framework for the theoretical predictions of RPIF is obtained [6].

In this work we will analyze the case of two particle beams, immersed in an inhomogeneous 
gravitational field, and consider the continuous monitoring of their positions. 
The idea here is to obtain a more realistic testing framework for RPIF, in which 
an experiment similar to Colella, Overhauser, and Werner (COW) case [1] could be performed. 
This more general situation could also shed some light on the controversy around 
the validity, at quantum level, of the equivalence principle [7], and also 
on the conceptual relation between quantum measurement theory and the weak equivalence principle [8].
\bigskip

\section{Measurements and Inhomogeneous Gravity.}
\bigskip
 
As has already been mentioned, in the attempts to solve the quantum measurement problem we may find RPIF [2]. 
This formalism explains a continuous quantum measurement with the introduction of a restriction on 
the integration domain of the corresponding path integral. 
This last condition can also be reformulated in terms 
of a weight functional that has to be considered in the path integral. 

Let us explain this point a little bit better, and suppose that we have a particle which shows one--dimensional movement. 
The amplitude $A(q'', q')$ for this particle to move from the point $q'$ to the point $q''$ is called propagator. 
It is given by Feynman [9] 

\begin{equation}
A(q'', q') = \int d[q]exp({i\over \hbar}S[q]),
\end{equation}

\noindent here we must integrate over all the possible trajectories $q(t)$, and $S[q]$ is the action of the system, which is 
defined as

\begin{equation}
S[q] = \int_{t'}^{t''}dtL(q, \dot{q}).
\end{equation}

Let us now suppose that we perform a continuous measurement of the position of this particle, 
such that we obtain as result of this measurement process a certain output $a(t)$. In other words, the measurement process gives the value $a(t)$ 
for the \-coor\-di\-na\-te $q(t)$ at each time $t$, and this output has associated a certain error $\Delta a$, which is determined by the 
experimental resolution of the measuring device. The amplitude $A_{[a]}(q'', q')$ can be now thought of as a probability amplitude for the continuous measurement process to give the result $a(t)$. 
Taking the square modulus of this amplitude allows us to find the probability density for different measurement outputs.

Clearly, the integration in the Feynman path--integral should be restricted to those trajectories that match with the experimental output. 
RPIF says that this condition can be introduced by means of a weight functional $\omega_a[q]$ [2]. 
This means that expression (1) 
becomes now under a continuous measurement process

\begin{equation}
A_a = \int d[q]\omega_a[q]exp(iS[q]).
\end{equation}

The more probable the trajectory $[q]$ is, according to the output $a$, the bigger that $\omega_a[q]$ becomes [2]. This means that the value of $\omega_a[q]$ is approximately one for all trajectories $[q]$ that agree with the measurement 
output $a$, and it is almost 0 for those that do not match with the result of the experiment. 
Clearly, the weight functional contains 
all the information about the interaction between measuring device and measured system. 

Consider the case of a particle with mass $m$ located in the Earth's 
gravitational field. Then its Lagrangian has the form
 
{\setlength\arraycolsep{2pt}\begin{eqnarray}
L = {\vec {P}^2\over2m} + {GMm\over r}.
\end{eqnarray}
\bigskip

Let us now write $r = R + l$, where $R$ denotes the Earth's radius and $l$ 
the distance above the Earth's surface. Under the condition $R>>l$ we may approximate 
the Lagrangian, up to second order in $l$, as follows
\bigskip

{\setlength\arraycolsep{2pt}\begin{eqnarray}
L = {\vec {P}^2\over2m} + {GMm\over R}\Bigl(1 - {l\over R} + {l^2\over R^2}\Bigr).
\end{eqnarray}
\bigskip

Now consider a quantum particle, whose Lagrangian is given by (5), then its 
propagator, if the particle goes from point $P$ to point $Q$, is

{\setlength\arraycolsep{2pt}\begin{eqnarray}
U(Q,\tau'';P, \tau') = \Bigl({m\over 2\pi i\hbar T}\Bigr)
exp\{{im\over 2\hbar T}[(x_Q - x_P)^2 + (y_Q - y_P)^2]\}\nonumber\\
\times\int d[l(t)]exp\{{i\over\hbar}\int_{\tau'}^{\tau''}
[{m\over 2}\dot{l}^2 + {GMm\over R} - {GMm\over R^2}l + {GMm\over R^3}l^2]dt\},
\end{eqnarray}}
\bigskip

here $\sqrt{(x_Q - x_P)^2 + (y_Q - y_P)^2}$ denotes the projection on the Earth's 
surface of the distance between $Q$ and $P$, and $T = \tau'' - \tau'$.

Our path integral may be rewritten as follows

{\setlength\arraycolsep{2pt}\begin{eqnarray}
\int d[l(t)]exp\{{i\over\hbar}\int_{\tau'}^{\tau''}
[{m\over 2}\dot{l}^2 + {GMm\over R} - {GMm\over R^2}l + {GMm\over R^3}l^2]dt\} =\nonumber\\
exp\{i{GMm\over\hbar R}T\}\int d[l(t)]exp\{{i\over\hbar}\int_{\tau'}^{\tau''}
[{m\over 2}\dot{l}^2 + F(t)l - {m\over 2}w^2l^2]dt\},
\end{eqnarray}}
\bigskip

being $F(t) = -{GMm\over R^2}$ and $w^2 = -{2GM\over R^3}$. Therefore the 
problem reduces to the calculation of the path integral of a driven harmonic oscilator, which 
has imaginary frequency $w = i\sqrt{{2GM\over R^3}} = i\Omega$ ($\Omega \in \Re$), 
and we already know the result of this evaluation [10].
\bigskip

In our case we have

{\setlength\arraycolsep{2pt}\begin{eqnarray}
\int d[l(t)]exp\{{i\over\hbar}\int_{\tau'}^{\tau''}
[{m\over 2}\dot{l}^2 + F(t)l - {m\over 2}w^2l^2]dt\} =\nonumber\\
\sqrt{{m\Omega\over 2\pi i\hbar \sinh(\Omega T)}}
exp\Bigl({im\Omega\over 2\hbar \sinh(\Omega T)}\{(l^2_Q + l^2_P)
\cosh(\Omega T) \nonumber\\
- 2l_Ql_P + R[1 - \cosh(\Omega T)]\Bigl[l_Q + l_P - {R\over 2}\Bigr] 
- \sqrt{{GMR\over 8}}T\sinh(\Omega T)\}\Bigr).
\end{eqnarray}}
\bigskip

This allows us to rewrite the propagator

{\setlength\arraycolsep{2pt}\begin{eqnarray}
U(Q,\tau'';P, \tau') = \Bigl({m\over 2\pi i\hbar T}\Bigr)^{3\over2}\sqrt{{\Omega T\over \sinh(\Omega T)}}
exp\{{im\over 2\hbar T}[(x_Q - x_P)^2 + (y_Q - y_P)^2]\}\nonumber\\
\times exp\{i{GMm\over\hbar R}T\}exp\Bigl({im\Omega\over 2\hbar \sinh(\Omega T)}\{(l^2_Q + l^2_P)
\cosh(\Omega T) \nonumber\\
- 2l_Ql_P + R[1 - \cosh(\Omega T)]\Bigl[l_Q + l_P - {R\over 2}\Bigr] 
- \sqrt{{GMR\over 8}}T\sinh(\Omega T)\}\Bigr).
\end{eqnarray}}
\bigskip

 Let us now introduce a measuring process, namely we will monitor continuously the 
 l--coordinate of the particle. Then expression (6) becomes now 
 
{\setlength\arraycolsep{2pt}\begin{eqnarray}
U_{[\alpha(t)]}(Q,\tau'';P, \tau') = \Bigl({m\over 2\pi i\hbar T}\Bigr)
exp\{{im\over 2\hbar T}[(x_Q - x_P)^2 + (y_Q - y_P)^2]\}\nonumber\\
\times exp\{i{GMm\over\hbar R}T\}\int d[l(t)]w_{[\alpha(t)]}[l(t)]exp\{{i\over\hbar}\int_{\tau'}^{\tau''}
[{m\over 2}\dot{l}^2 + F(t)l - {m\over 2}w^2l^2]dt\}.
\end{eqnarray}}
\bigskip

The modulus square of (10) gives the probability of obtaining 
as measurement output (for the $l$--coordinate) function $\alpha(t)$. 
The weight functional $w_{[\alpha(t)]}[l(t)]$ contains the information concerning the measurement, and 
is determined by the experimental construction [2].

At this point, in order to obtain theoretical predictions, we must choose a 
par\-ti\-cu\-lar expression 
for $w_{[\alpha(t)]}[l(t)]$. 
We know that the results coming from a Heaveside weight functional [11] and those 
coming from a gaussian one [12] coincide up to the order of magnitude. 
These last remarks allow us to consider a gaussian weight functional as an 
approximation of the correct expression.
But a sounder justification of this choice comes from the fact that there 
are measuring processes in which the weight functional has precisely a gaussian form [13]. 
In consequence we could think about a measuring device whose weight functional is very close to a gaussian behavior.

Therefore we may now choose as our weight functional the following expression 

\begin{equation}
\omega_{[\alpha(t)]}[l(t)] = exp\{-{2\over T\Delta \alpha^2}\int _{\tau '}^{\tau ''}[l(t) - \alpha(t)]^2dt\},
\end{equation}
\bigskip

here $\Delta \alpha$ represents the error in our measurement.

Hence, with the introduction of a continuous quantum measurement the new pro\-pa\-ga\-tor is

{\setlength\arraycolsep{2pt}\begin{eqnarray}
U_{[\alpha(t)]}(Q,\tau'';P, \tau') = \Bigl({m\over 2\pi i\hbar T}\Bigr)
exp\{{im\over 2\hbar T}[(x_Q - x_P)^2 + (y_Q - y_P)^2]\}\nonumber\\
\times exp\{i{GMm\over\hbar R}T\}exp\{-{2\over T\Delta \alpha^2}\int _{\tau '}^{\tau ''}[l(t) - \alpha(t)]^2dt\}\nonumber\\
\times \int d[l(t)]exp\{{i\over\hbar}\int_{\tau'}^{\tau''}
[{m\over 2}\dot{l}^2 + F(t)l - {m\over 2}w^2l^2]dt\}.
\end{eqnarray}}
\bigskip

Once again we find the case of a driven harmonic oscillator, but now the 
involved forces and frequencies have a nonvanishing imaginary part. Besides, the driving 
force term is not constant, this nontrivial time dependence arises from the 
presence of the term ${4i\hbar\alpha(t)\over T\Delta\alpha^2}$ (see below).

(12) may be rewritten as follows

{\setlength\arraycolsep{2pt}\begin{eqnarray}
U_{[\alpha_(t)]}(Q,\tau'';P, \tau') = \Bigl({m\over 2\pi i\hbar T}\Bigr)
exp\{{im\over 2\hbar T}[(x_Q - x_P)^2 + (y_Q - y_P)^2]\} \nonumber\\
\times exp\{i{GMm\over R\hbar}T\}exp\{-{2\over T\Delta \alpha^2}\int_{\tau'}^{\tau''}\alpha^2(t)dt\}\nonumber\\
\times\int d[l(t)]exp\{{i\over\hbar}\int_{\tau'}^{\tau''}[{m\over 2}\dot{l}^2 + 
\hat{F}(t)l -{m\over 2}\hat{\omega}^2l^2]dt\}.
\end{eqnarray}}
\bigskip
 
In this last expression we have introduced the following definitions 
$\hat{F}(t) = -{GMm\over R^2} - {4i\hbar\alpha(t)\over T\Delta \alpha^2}$ and 
$\hat{\omega}^2 =  -{2GM\over R^3} - {4i\hbar\over mT\Delta \alpha^2}$. 
\bigskip

After a lengthy calculation we obtain the propagator for a particle whose $l$--coordinate 
has been continuously monitored.

{\setlength\arraycolsep{2pt}\begin{eqnarray}
U_{[\alpha_(t)]}(Q,\tau'';P, \tau') = \Bigl({m\over 2\pi i\hbar T}\Bigr)^{3\over2}
\sqrt{{\hat{\Omega} T\over \sinh(\hat{\Omega} T)}}\nonumber\\
exp\{{im\over 2\hbar T}[(x_Q - x_P)^2 + (y_Q - y_P)^2]\}exp\{i{GMm\over R\hbar}T\}\nonumber\\
\times exp\{-{2\over T\Delta \alpha^2}\int_{\tau'}^{\tau''}\alpha^2(t)dt\}
exp\Bigl({im\hat{\Omega}\over 2\hbar \sinh(\hat{\Omega} T)}\{(l^2_Q + l^2_P)\cosh(\hat{\Omega} T) \nonumber\\
- 2l_Ql_P + {R\over 1 + \gamma}[1 - \cosh(\hat{\Omega} T)]\Bigl[l_Q + l_P - {R\over 2(1 + \gamma)}\Bigr] \nonumber\\
- \sqrt{{GMR\over 8(1 + \gamma)^3}}T\sinh(\hat{\Omega} T)\nonumber\\
- {8i\hbar\over T\Delta\alpha^2m}\sqrt{{R^3\over 2GM(1 + \gamma)}}
[l_QF^{(1)}(\tau'', \tau') + l_PF^{(2)}(\tau'', \tau')] \nonumber\\
+ ({4\hbar\over T\Delta \alpha^2m})^2{R^3\over GM(1 + \gamma)}F^{(3)}(\tau'', \tau') \nonumber\\
+ {4i\hbar R\over T\Delta \alpha^2m}\sqrt{{R^3\over 2GM(1 + \gamma)}}
[F^{(2)}(\tau'', \tau') - F^{(4)}(\tau'', \tau') \nonumber\\
- \int_{\tau'}^{\tau''}F^{(1)}(\tau, \tau')\sinh(\hat{\Omega} (\tau'' - \tau))d\tau] \}\Bigr).
\end{eqnarray}}
\bigskip

In this last expression we have $\hat{\Omega} =  \sqrt{{2GM\over R^3}(1 + \gamma)}$, $\gamma = {2i\hbar R^3\over GMmT\Delta \alpha^2}$, 
$F^{(2)}(\tau'', \tau') = \int_{\tau'}^{\tau''}\alpha(\tau)\sinh(\hat{\Omega} [\tau'' - \tau])d\tau$, 
$F^{(1)}(\tau'', \tau') = \int_{\tau'}^{\tau''}\alpha(\tau)\sinh(\hat{\Omega} [\tau - \tau'])d\tau$, 
$F^{(3)}(\tau'', \tau') = \int_{\tau'}^{\tau''}d\tau\int_{\tau'}^{\tau}ds\alpha(\tau)\alpha(s)
\sinh(\hat{\Omega} [\tau'' - \tau])\sinh(\hat{\Omega} [s - \tau'])$, and finally we have also introduced $F^{(4)}(\tau'', \tau') = \int_{\tau'}^{\tau''}\alpha(\tau)\sinh(\hat{\Omega} [\tau'' - \tau])
\cosh(\hat{\Omega} [\tau - \tau'])d\tau$. 
\bigskip
 
Let us now suppose that we have two particles, which start at point $P$ and are 
detected at point $Q$. The $l$--coordinate of both particles is going to be continuously 
monitored, and in this measuring process we use measuring devices with different 
experimental resolutions, in other words, we have not only two different trajectories, 
$\alpha(t)$ and $\beta(t)$, as measurement outputs, but we also have 
$\Delta \alpha \not = \Delta \beta$.
  
Under these conditions the emerging interference pattern is obtained using two expressions 
like that given in (14).

The resulting interference pattern can be written as follows

{\setlength\arraycolsep{2pt}\begin{eqnarray}
I = I_1 + I_2 + I_3 + I_4 + I_5.
\end{eqnarray}}

Here we have that 

{\setlength\arraycolsep{2pt}\begin{eqnarray}
I_1 = {m\over 2\hbar}(l^2_Q + l^2_P)\Bigl[{\tilde{\Omega}\sin(2\check{\Omega}T) 
- \check{\Omega}\sinh(2\tilde{\Omega}T)\over \cosh(2\tilde{\Omega}T) - \cos(2\check{\Omega}T)} - 
{\tilde{\Gamma}\sin(2\check{\Gamma}T) 
- \check{\Gamma}\sinh(2\tilde{\Gamma}T)\over \cosh(2\tilde{\Gamma}T) - \cos(2\check{\Gamma}T)}\Bigr],
\end{eqnarray}}

{\setlength\arraycolsep{2pt}\begin{eqnarray}
I_2 = -2l_Ql_P{m\over \hbar}\Bigl[{-\check{\Omega}\sinh(\tilde{\Omega}T)\cos(\check{\Omega}T) + 
\tilde{\Omega}\cosh(\tilde{\Omega}T)\sin(\check{\Omega}T)\over \cosh(2\tilde{\Omega}T) - \cos(2\check{\Omega}T)} \nonumber\\
- {-\check{\Gamma}\sinh(\tilde{\Gamma}T)\cos(\check{\Gamma}T) + 
\tilde{\Gamma}\cosh(\tilde{\Gamma}T)\sin(\check{\Gamma}T)\over \cosh(2\tilde{\Gamma}T) - \cos(2\check{\Gamma}T)}\Bigr] \nonumber\\
+ {mT\over 2\hbar}\sqrt{{GMR\over 8}}\Bigl[[1 + \tilde{\gamma}^2]^{-3/4}
[-\check{\Omega}\cos({3\over 2}\arctan(\tilde{\gamma})) + \tilde{\Omega}\sin({3\over 2}\arctan(\tilde{\gamma}))]\nonumber\\
- [1 + \tilde{\eta}^2]^{-3/4}
[-\check{\Gamma}\cos(-{3\over 2}\arctan(\tilde{\eta})) + \tilde{\Gamma}\sin(-{3\over 2}\arctan(\tilde{\eta}))] \Bigr],
\end{eqnarray}}

{\setlength\arraycolsep{2pt}\begin{eqnarray}
I_3 = {m\over \hbar}\Bigl[{-\check{\Omega}\sinh(\tilde{\Omega}T)\cos(\check{\Omega}T) + 
\tilde{\Omega}\cosh(\tilde{\Omega}T)\sin(\check{\Omega}T)\over \cosh(2\tilde{\Omega}T) - \cos(2\check{\Omega}T)} \nonumber\\
\times\Bigl[[1 - \cosh(\tilde{\Omega}T)\cos(\check{\Omega}T)]{R\over 1 + \tilde{\gamma}^2}
[l_Q + l_P - {R(1 - \tilde{\gamma}^2)\over 2(1 + \tilde{\gamma}^2)}]\nonumber\\
+ {R\tilde{\gamma}\over 1 + \tilde{\gamma}^2}[{R\over 1 + \tilde{\gamma}^2} - (l_Q + l_P)]\sinh(\tilde{\Omega}T)\sin(\check{\Omega}T)\Bigr]\nonumber\\
-{m\over \hbar}\Bigl[{\check{\Omega}\cosh(\tilde{\Omega}T)\sin(\check{\Omega}T) + 
\tilde{\Omega}\sinh(\tilde{\Omega}T)\cos(\check{\Omega}T)\over \cosh(2\tilde{\Omega}T) - \cos(2\check{\Omega}T)} \nonumber\\
\times\Bigl[[1 - \cosh(\tilde{\Omega}T)\cos(\check{\Omega}T)]{R\tilde{\gamma}\over 1 + \tilde{\gamma}^2}
[{R\over 1 + \tilde{\gamma}^2} - (l_Q + l_P)]\nonumber\\
- {R\over 1 + \tilde{\gamma}^2}[l_Q + l_P - {R(1 - \tilde{\gamma}^2)\over 2(1 + \tilde{\gamma}^2)}]\sinh(\tilde{\Omega}T)\sin(\check{\Omega}T)\Bigr]\nonumber\\
-{m\over \hbar}\Bigl[{-\check{\Gamma}\sinh(\tilde{\Gamma}T)\cos(\check{\Gamma}T) + 
\tilde{\Gamma}\cosh(\tilde{\Gamma}T)\sin(\check{\Gamma}T)\over \cosh(2\tilde{\Gamma}T) - \cos(2\check{\Gamma}T)} \nonumber\\
\times\Bigl[[1 - \cosh(\tilde{\Gamma}T)\cos(\check{\Gamma}T)]{R\over 1 + \tilde{\eta}^2}
[l_Q + l_P - {R(1 - \tilde{\eta}^2)\over 2(1 + \tilde{\eta}^2)}]\nonumber\\
+ {R\tilde{\eta}\over 1 + \tilde{\eta}^2}[l_Q + l_P - {R\over 1 + \tilde{\eta}^2}]\sinh(\tilde{\Gamma}T)\sin(\check{\Gamma}T)\Bigr]\nonumber\\
+{m\over \hbar}\Bigl[{\check{\Gamma}\cosh(\tilde{\Gamma}T)\sin(\check{\Gamma}T) + 
\tilde{\Gamma}\sinh(\tilde{\Gamma}T)\cos(\check{\Gamma}T)\over \cosh(2\tilde{\Gamma}T) - \cos(2\check{\Gamma}T)} \nonumber\\
\times\Bigl[[1 - \cosh(\tilde{\Gamma}T)\cos(\check{\Gamma}T)]{R\tilde{\eta}\over 1 + \tilde{\eta}^2}
[l_Q + l_P - {R\over 1 + \tilde{\eta}^2}]\nonumber\\
- {R\over 1 + \tilde{\eta}^2}[l_Q + l_P - {R(1 - \tilde{\eta}^2)\over 2(1 + \tilde{\eta}^2)}]\sinh(\tilde{\Gamma}T)\sin(\check{\Gamma}T)\Bigr].
\end{eqnarray}}

Here $\tilde{\Omega} = \sqrt{{2GM\over R^3}}[1 + \tilde{\gamma}^2]^{1/4}
\cos[{1\over 2}\arctan(\tilde{\gamma})]$, $\check{\Omega} = \sqrt{{2GM\over R^3}}[1 + \tilde{\gamma}^2]^{1/4}
\sin[{1\over 2}\arctan(\tilde{\gamma})]$, with $\tilde{\gamma} = {2\hbar R^3\over GMmT\Delta \alpha^2}$. 
Also $\tilde{\Gamma} = \sqrt{{2GM\over R^3}}[1 + \tilde{\eta}^2]^{1/4}
\cos[-{1\over 2}\arctan(\tilde{\eta})]$, and  
$\check{\Gamma} = \sqrt{{2GM\over R^3}}[1 + \tilde{\eta}^2]^{1/4}
\sin[-{1\over 2}\arctan(\tilde{\eta})]$, with $\tilde{\eta} = {2\hbar R^3\over GMmT\Delta \beta^2}$.
\bigskip

The fourth term is

{\setlength\arraycolsep{2pt}\begin{eqnarray}
I_4 = {8\over T\Delta\alpha^2}\Bigl[{\check{\Omega}\cosh(\tilde{\Omega}T)\sin(\check{\Omega}T) + 
\tilde{\Omega}\sinh(\tilde{\Omega}T)\cos(\check{\Omega}T)\over \cosh(2\tilde{\Omega}T) - \cos(2\check{\Omega}T)}\Bigr]\nonumber\\
\times\Bigl[{\tilde{\Omega}\over\tilde{\Omega}^2 + \check{\Omega}^2}[l_Qf^{(1)}(\tau'', \tau') + l_P\tilde{f}^{(1)}(\tau'', \tau')] \nonumber\\
+ {\check{\Omega}\over\tilde{\Omega}^2 + \check{\Omega}^2}[l_Qf^{(2)}(\tau'', \tau') + l_P\tilde{f}^{(2)}(\tau'', \tau')]\Bigr]\nonumber\\
+ {8\over T\Delta\alpha^2}\Bigl[{-\check{\Omega}\sinh(\tilde{\Omega}T)\cos(\check{\Omega}T) + 
\tilde{\Omega}\cosh(\tilde{\Omega}T)\sin(\check{\Omega}T)\over \cosh(2\tilde{\Omega}T) - \cos(2\check{\Omega}T)}\Bigr]\nonumber\\
\times\Bigl[{\check{\Omega}\over\tilde{\Omega}^2 + \check{\Omega}^2}[l_Qf^{(1)}(\tau'', \tau') + l_P\tilde{f}^{(1)}(\tau'', \tau')] \nonumber\\
- {\tilde{\Omega}\over\tilde{\Omega}^2 + \check{\Omega}^2}[l_Qf^{(2)}(\tau'', \tau') + l_P\tilde{f}^{(2)}(\tau'', \tau')]\Bigr]\nonumber\\
-{8\over T\Delta\beta^2}\Bigl[{\check{\Gamma}\cosh(\tilde{\Gamma}T)\sin(\check{\Gamma}T) + 
\tilde{\Gamma}\sinh(\tilde{\Gamma}T)\cos(\check{\Gamma}T)\over \cosh(2\tilde{\Gamma}T) - \cos(2\check{\Gamma}T)}\Bigr]\nonumber\\
\times\Bigl[{\tilde{\Gamma}\over\tilde{\Gamma}^2 + \check{\Gamma}^2}[l_Qf^{(3)}(\tau'', \tau') + l_P\tilde{f}^{(3)}(\tau'', \tau')] \nonumber\\
+ {\check{\Gamma}\over\tilde{\Gamma}^2 + \check{\Gamma}^2}[l_Qf^{(4)}(\tau'', \tau') + l_P\tilde{f}^{(4)}(\tau'', \tau')]\Bigr]\nonumber\\
- {8\over T\Delta\beta^2}\Bigl[{-\check{\Gamma}\sinh(\tilde{\Gamma}T)\cos(\check{\Gamma}T) + 
\tilde{\Gamma}\cosh(\tilde{\Gamma}T)\sin(\check{\Gamma}T)\over \cosh(2\tilde{\Gamma}T) - \cos(2\check{\Gamma}T)}\Bigr]\nonumber\\
\times\Bigl[-{\check{\Gamma}\over\tilde{\Gamma}^2 + \check{\Gamma}^2}[l_Qf^{(3)}(\tau'', \tau') + l_P\tilde{f}^{(3)}(\tau'', \tau')] \nonumber\\
+ {\tilde{\Gamma}\over\tilde{\Gamma}^2 + \check{\Gamma}^2}[l_Qf^{(4)}(\tau'', \tau') + l_P\tilde{f}^{(4)}(\tau'', \tau')]\Bigr].
\end{eqnarray}}

Here we have defined $f^{(1)}(\tau'', \tau') = \int_{\tau'}^{\tau''}\alpha(\tau)\sinh(\tilde{\Omega}[\tau - \tau'])cos(\check{\Omega}[\tau - \tau'])d\tau$. Also  
$\tilde{f}^{(1)}(\tau'', \tau') = \int_{\tau'}^{\tau''}\alpha(\tau)\sinh(\tilde{\Omega}[\tau'' - \tau])cos(\check{\Omega}[\tau'' - \tau])d\tau$. 
Additionally we have introduced $f^{(2)}(\tau'', \tau') = \int_{\tau'}^{\tau''}\alpha(\tau)\cosh(\tilde{\Omega}[\tau - \tau'])sin(\check{\Omega}[\tau - \tau'])d\tau$. Similarly 
$\tilde{f}^{(2)}(\tau'', \tau') = \int_{\tau'}^{\tau''}\alpha(\tau)\cosh(\tilde{\Omega}[\tau'' - \tau])sin(\check{\Omega}[\tau'' - \tau])d\tau$.

Concerning the remaining definitions we have $f^{(3)}(\tau'', \tau') = \int_{\tau'}^{\tau''}\beta(\tau)\sinh(\tilde{\Gamma}[\tau - \tau'])cos(\check{\Gamma}[\tau - \tau'])d\tau$. Also  
$\tilde{f}^{(3)}(\tau'', \tau') = \int_{\tau'}^{\tau''}\beta(\tau)\sinh(\tilde{\Gamma}[\tau'' - \tau])cos(\check{\Gamma}[\tau'' - \tau])d\tau$. 
In addition we have employed $f^{(4)}(\tau'', \tau') = \int_{\tau'}^{\tau''}\beta(\tau)\cosh(\tilde{\Gamma}[\tau - \tau'])sin(\check{\Gamma}[\tau - \tau'])d\tau$. Finally, 
$\tilde{f}^{(4)}(\tau'', \tau') = \int_{\tau'}^{\tau''}\beta(\tau)\cosh(\tilde{\Gamma}[\tau'' - \tau])sin(\check{\Gamma}[\tau'' - \tau])d\tau$.
\bigskip

The last contribution to the interference term reads

{\setlength\arraycolsep{2pt}\begin{eqnarray}
I_5 = Re\Bigl[{im\hat{\Omega}\over 2\hbar \sinh(\hat{\Omega} T)}
({4\hbar\over T\Delta \alpha^2m})^2{R^3\over GM(1 + i\tilde{\gamma})}F^{(3)}(\tau'', \tau')\Bigr]\nonumber\\
{-im\hat{\Gamma}\over 2\hbar \sinh(\hat{\Gamma} T)}({4\hbar\over T\Delta \beta^2m})^2{R^3\over GM(1 - i\tilde{\eta})}\tilde{F}^{(3)}(\tau'', \tau')\nonumber\\
+ Re\Bigl[-{2R\hat{\Omega}\over T\Delta\alpha^2 \sinh(\hat{\Omega} T)}\Bigr]\sqrt{{R^3\over 2GM(1 + i\tilde{\gamma})}}
[F^{(2)}(\tau'', \tau') - F^{(4)}(\tau'', \tau') \nonumber\\
- \int_{\tau'}^{\tau''}F^{(1)}(\tau, \tau')\sinh(\hat{\Omega} (\tau'' - \tau))d\tau + 
{2R\hat{\Gamma}\over T\Delta\beta^2 \sinh(\hat{\Gamma} T)}\Bigr]\nonumber\\
\times\sqrt{{R^3\over 2GM(1 - i\tilde{\eta})}}[\tilde{F}^{(2)}(\tau'', \tau') - \tilde{F}^{(4)}(\tau'', \tau') \nonumber\\
- \int_{\tau'}^{\tau''}\tilde{F}^{(1)}(\tau, \tau')\sinh(\hat{\Gamma} (\tau'' - \tau))d\tau].
\end{eqnarray}}

Here $\tilde{F}^{(2)}(\tau'', \tau') = \int_{\tau'}^{\tau''}\beta(\tau)\sinh(\hat{\Gamma} [\tau'' - \tau])d\tau$, 
$\tilde{F}^{(1)}(\tau'', \tau') = \int_{\tau'}^{\tau''}\beta(\tau)\sinh(\hat{\Gamma} [\tau - \tau'])d\tau$, 
$\tilde{F}^{(3)}(\tau'', \tau') = \int_{\tau'}^{\tau''}d\tau\int_{\tau'}^{\tau}ds\beta(\tau)\beta(s)
\sinh(\hat{\Gamma} [\tau'' - \tau])\sinh(\hat{\Gamma} [s - \tau'])$, and finally we have also introduced $\tilde{F}^{(4)}(\tau'', \tau') = \int_{\tau'}^{\tau''}\beta(\tau)\sinh(\hat{\Gamma} [\tau'' - \tau])
\cosh(\hat{\Gamma} [\tau - \tau'])d\tau$, being $\hat{\Gamma} =  \tilde{\Gamma} + i\check{\Gamma}$. 

\bigskip

\section{Conclusions.}
\bigskip

Our results are a generalization (in the present work we have an inhomogeneous 
gravitational field) of expression (14) of reference [6] (which was deduced for the case of a homogeneous field). 

Let us now analyze the effects that an inhomogeneous gravitational field has upon 
the emerging interference pattern. In order to do this, as a first step, we compare 
expressions (16), (17), and (18) with the corresponding ones in the case of a homogeneous field, 
expressions (15) and (16) in [6]. Clearly, if we consider $I_1$, $I_2$, and $I_3$ as functions of the resolutions 
of the measuring devices, then we may see that the introduction of an inhomogeneous field renders 
an interference pattern that changes more rapidly (as function of these resolutions) 
than in the case of a homogeneous situation, this feature could be, in principle, detected. 

We may understand this new characteristic noting that in the case of a homogeneous field 
the ``frequency'' of the complex harmonic oscillator is purely imaginary ($w = \sqrt{-i{4\hbar\over mT\Delta \alpha^2}}$). 
The case of an inhomogeneous gra\-vi\-ta\-tional field endows the co\-rres\-pond\-ing ``frequency'' 
with a non--vanishing real part ($\hat{\omega} =  \sqrt{-{2GM\over R^3} - {4i\hbar\over mT\Delta \alpha^2}}$). 
This means that the phase of each particle has a new dependence, and in consequence 
the phase shift, between the beams, that at the end appears, must contain this new information. 
In other words, the difference in the interference pattern between the present work and [6] stems from the presence of a non--vanishing real term in the ``frequency'' 
of the oscillator, expression (13). 

The introduction of an inhomogeneous gravitational field does not imply the pre\-sen\-ce 
of a new term in the driving force for our oscillator. Indeed, if we take a look at the corresponding 
expression for the inhomogeneous situation, $\hat{F}(t) = -{GMm\over R^2} - {4i\hbar\alpha(t)\over T\Delta \alpha^2}$, 
and compare it with the force in [6], then, remembering that $g = {GM\over R^2}$, 
we find that they are the same. 
This last remark also shows us that in the homogeneous case the information concerning 
the geometry of the source of the gravitational field comes from the driving force 
term in the corresponding oscillator, and appears only in expressions (17), (18), and (19). 
The two first contributions to the interference term ((15) and (16) in [6]) do not contain any information about 
the geometry of the source. 

But, in this new case case, we see that now all terms do contain this information, for instance, 
$\tilde{\Omega}$ and $\check{\Omega}$ are now functions of ${2\hbar R^3\over GMmT\Delta \alpha^2}$, i.e., the radius of the Earth appears in this expression 
(in the homogeneous case the respective parameter is a function of  
${2\pi\hbar T\over m\Delta \alpha^2}$). 
This is a trivial remark, nevertheless renders an important di\-ffe\-ren\-ce between the predictions of the homogeneous and inhomogeneous cases. Indeed, if we restrict ourselves at this point 
to expressions $I_1$, $I_2$, and $I_3$, then we may see that if we carry out the experiment 
not on the Earth's surface, but at a certain height $\tilde{R}$ above this surface, 
the inhomogeneous case predicts a change in the interference pattern (under this 
new condition we would have that $\tilde{\Omega}$ and $\check{\Omega}$ are now functions of 
${2\hbar (R + \tilde{R})^3\over GMmT\Delta \alpha^2}$). This change in the interference 
pattern appears even if all the remaining involved parameters do not suffer any modification. 
In the homogeneous case, $I_1$ and $I_2$ do not suffer any modification under with change, 
because the ``frequency''  of the corresponding oscillator does not contain any information 
about the geometry of the source of the gravitational field.

Therefore, we may state that the difference in these two cases resides in the fact 
that in the extension that comprises the inhomogeneous field we are able to include 
in the ``frequency'' of the complex oscillator the effects of the geometry of the source of the gravitational field, something that is not possible in the first case.

Let us now compare our predictions with the measurement outputs of the COW experiment.
If we analyze COW, taking into account also an inhomogeneous field, 
then we find that the interference pattern is given by $I = cos\{-{gm^2Ll_b\Lambda\over\hbar^2}[1 -{l_b\over R}]\}$, 
here $\Lambda$ denotes the initial reduced wavelength of the packets, $L$ the horizontal 
separation between starting point and detection point, and finally $l_b$ is the vertical 
separation between these two points. 
The term ${l_b\over R}$ does not appear in COW, but it can be deduced if we include 
an inhomogeneous field [14]. The dependence of the cosine function upon the distance 
to the center of the Earth is given by ${GMm^2Ll_b\Lambda\over\hbar^2R^2}[1 -{l_b\over R}]$. 
Clearly, expressions (16), (17), and (18) contain a very di\-ffe\-rent dependence on $R$. 
We have two very different mathematical dependences on the geometry of the source. 
In principle, this difference could be detected. 

Concerning the feasibility of an experiment with these characteristics, for the time being, it seems that the present technology can not cope with the 
experimental difficulties that this proposal implies [15]. Nevertheless, in order 
to have a very rough estimation of the possible order of magnitude of the emerging effects 
that we could expect, let us at this point suppose that this kind of experiments 
could be carried out with the same accuracy that nowadays exist in the case of Paul traps [16], 
in which an individual ion is trapped employing a high--frequency electric quadrupole. 
If we take a look at expressions (16), (17), and (18) we will see that one of the points that could be important 
in this feasibility is the order of magnitude of $\check{\Gamma}$, $\check{\Omega}$, $\tilde{\Gamma}$, and also of 
$\tilde{\Omega}$. The bound for these parameters is going to be determined 
by $\sqrt{{2GM\over R^3}}[1 + \tilde{\gamma}^2]^{1/4}$ and $\sqrt{{2GM\over R^3}}[1 + \tilde{\eta}^2]^{1/4}$. 
Here the resolution of the measuring device is the most important factor ($\tilde{\eta} = {2\hbar R^3\over GMmT\Delta \beta^2}$ and 
$\tilde{\gamma} = {2\hbar R^3\over GMmT\Delta \alpha^2}$). If we assume the resolution of a typical Paul trap [17], $\Delta \alpha \sim 2\mu$m, 
then we may see that $\sqrt{\tilde{\gamma}} \sim 10^{5}$ (here we consider, as in the COW experiment, thermal neutrons), in consequence the bound is $ \sim 10^{2}s^{-1}$. 
Though this is a very rough estimation of $\tilde{\gamma}$ (also of $\tilde{\eta}$) 
it seems to claim that if the development of the current technology associated with Paul and Penning traps 
could, in a future, measure continuously the position of individual atoms, then this 
kind of proposals could render non--trivial measurement outputs.

It is also clear that even if we had the case $\alpha(t) = \beta(t)$, the condition $\Delta\alpha \not = \Delta\beta$ 
would render a non--vanishing interference pattern. 
This resolution--induced in\-ter\-fe\-ren\-ce is a new feature in this context 
and renders an a\-ddi\-tio\-nal, and also less restricted, testing framework for the theoretical predictions 
of RPIF. 
Remembering the COW experiment [1] (where interference vanishes if the particles 
follow the same trajectory) we may conclude that the introduction of a 
measuring process implies the appearance, in the interference term, of the mass parameter, 
even if the condition $\alpha(t) = \beta(t)$ is fulfilled. 
This last result is, qualitatively, very similar to the behavior of the 
master equation of a freely falling particle, namely, the presence of a 
measurement coupling implies the appearance of the mass of the corresponding particles 
as a crucial factor in the dynamics of the system [8].

We shall not forget that our main result 
was deduced with the introduction of one approximation ($l<<R$), and therefore can not have general validity.  
The general result, which will be published elsewhere, is important, not only because it 
could allow the introduction of less restricted experimental proposals, but also because 
it could render a more profound understanding of the conceptual relation between 
quantum measurement and the weak equivalence principle [8]. At the same time it could also shed some 
light on the controversy around the validity, at quantum level, of the equivalence principle [7].

Clearly gravity at quantum level continues to be a not purely geometric effect, mass 
appears in the interference pattern (always in the combination $\hbar/m$). 

\bigskip

\Large{\bf Acknowledgments.}\normalsize
\bigskip

The author would like to thank A. Camacho--Galv\'an and A. A. Cuevas--Sosa for their 
help, and D.-E. Liebscher for the fruitful discussions on the subject. 
The hospitality of the Astrophysikalisches Institut Potsdam is also kindly acknowledged. 
This work was supported by CONACYT Posdoctoral Grant No. 983023.

\end{document}